\documentclass[aps,prb,preprint,groupedaddress,showpacs]{revtex4-1}
\usepackage{graphicx}
\begin{document}

\title{Resistivity tensor correlations in the mixed state of electron-doped superconductor  Nd$_{2-x}$Ce$_x$CuO$_{4+\delta}$}

\author{T.B.Charikova$^{\rm 1}$$^,$$^{\rm 2}$, N.G.Shelushinina$^{\rm 1}$, G.I.Harus$^{\rm 1}$, D.S.Petukhov$^{\rm 1}$, O.E.Petukhova$^{\rm 1}$,
A.A.Ivanov$^{\rm 3}$}

\thanks{}
\affiliation{$^{\rm 1}$M.N.Mikheev Institute of Metal Physics Ural Branch of RAS, Ekaterinburg, Russia}
\affiliation{$^{\rm 2}$ Ural Federal University, Ekaterinburg, Russia} 
\affiliation{$^{\rm 3}$ National Research Nuclear University MEPhI, Moscow, Russia}

\date{\today}

\begin{abstract}
The magnetic-field dependencies of the longitudinal and Hall resistance of the electron-doped compounds Nd$_{2-x}$Ce$_x$CuO$_{4+\delta}$ in underdoped region with $x$ = 0.14 and with varying degrees of disorder ($\delta$) were investigated. It was established experimentally that the correlation between the longitudinal electrical resistivity and the Hall resistivity can be analyzed on the basis of scaling relationships: $\rho_{xy}$($B$)$\sim$ $\rho^{\beta}_{xx}$($B$). For the totality of the investigated single-crystal films of Nd$_{2-x}$Ce$_x$CuO$_{4+\delta}$/SrTiO$_3$ the universal value $\beta$ = 1.5 $\pm$0.4 is found. The observed feature in the
electron-doped two-dimensional systems can be associated both with a displaying of  anisotropic $s$ - wave or  $d$-wave pairing symmetry and with a rather strong pinning due to an essential degree of disorder in the samples under study.

\end{abstract}

\pacs{74.72.Ek, 74.78.-w, 74.25.F-}

\maketitle

\section{Introduction}

The change of the Hall coefficient sign in the normal state with changing of the  dopant content and the sign inversion (sometimes double) in the mixed state of superconducting systems is currently a hot topic of research   \cite{Berg}$^-$\cite{Charikova1}. The study of the Hall effect, together with the study of electric and magnetoresistance \cite{Charikova2}$^-$\cite{Guarino}, and Nernst effect \cite{Tafti} is a key point in the theoretical and experimental analysis of the charge system of high-temperature superconductors (HTSC) and electron-doped high-temperature superconductors, in particular. Moreover, in the electron-doped HTSC two points are important in changing the concentration and type of carriers (electrons and holes): the contents of the dopant and the level of non-stoichiometric oxygen. The partaking of the two types of carriers in the superconducting (SC) phase and the formation of other phases with different order parameters, such as the antiferromagnetic (AF), the spin density wave (SDW) and charge density wave (CDW) ones, are actively discussed now \cite{Harshman}$^-$\cite{deSilva}. The proposed models and theories for the hole-doped high-temperature superconductors \cite{Dorsey1}$^-$\cite{Wang1} can not fully explain the features of the behavior of the tensor resistivity in the electron-doped superconducting compounds.

One of the most striking features of the vortex motion in the oxide superconductors is the behavior of the Hall resistivity which has attracted intense attention. Already in 1968 R.R.Hake \cite{Hake} during the mixed state Hall effect study of a II-nd type superconductor Ti-16 at $\%$ Mo have marked features of the field dependence of the Hall angle in a mixed state in comparison with the normal state of a superconductor.  In the most of the hole-doped high-Tc compounds it has been found that the correlation between the longitudinal and the Hall resistivity in the mixed region can be analyzed on the basis of the scaling relations $\rho_{xy}$($B$)$\sim$ $\rho^{\beta}_{xx}$($B$)  \cite{Luo}$^-$\cite{Kang5}. 

In particular,  for the first time observation of a puzzling  scaling behavior of the Hall versus longitudinal resistivities  has been reported by Luo et al.  \cite{Luo}  in the mixed state of epitaxial YBa$_2$Cu$_3$0$_7$ films: $\rho_{xy}$($T$) $\propto$ $\rho^{\beta}_{xx}$($T$) at fixed field, with $\beta$ = 1.7 $\pm$ 0.2.  Furthermore, a scaling relation between $\rho_{xy}$ and $\rho_{xx}$, $\rho_{xy}$ =  A$\rho^{\beta} _{xx}$, with $\beta$ $\simeq$ 2 has been observed for Bi$_2$Sr$_2$CaCu$_2$O$_8$ crystals \cite{Samilov1}  and Tl$_2$Ba$_2$Ca$_2$Cu$_3$O$_10$ films \cite{Budhani} . Other similar studies have reported $\beta$ = 1.5 $-$ 2.0 for YBa$_2$Cu$_3$O$_7$  crystals  \cite{Kang3}$^-$\cite{Gob} and HgBa$_2$CaCu$_2$O$_6$ films \cite{Kang4}. Even $\beta$ $\simeq$ 1 was reported for heavy-ion- irradiated HgBa$_2$CaCu$_2$O$_6$ thin films \cite{Kang5}. 

For the electron-doped compounds the Hall effect in a mixed state was investigated on single crystals of L$_{1.85}$Ce$_{0.15}$CuO$_ {4} $ (L = Nd, Sm) with an optimal doping ($x$ = 0.15) \cite{Cagigal95}. It was also observed scaling relations between longitudinal electrical resistivity and the Hall resistivity, but the exponent differs from the value of $\beta$ for the hole-doped compounds, and is equal to $\beta$ = 0.8 $\pm$ 0.2. 

Recently a scaling behavior between the Hall resistivity and the longitudinal resistivity, $\rho_{xy}$ =  A $\rho^{\beta} _{xx}$, with a universal power law ($\beta$ = 2.0 $\pm$ 0.1) has been observed in MgB$_2$ thin films for a wide magnetic field and temperature regions \cite{Kang6}$^,$\cite{Soon}. It was also reported about  a scaling law between the Hall  $\rho_{xy}$($B$) and the longitudinal $\rho_{xx}$($B$) resistivity in the superconducting transition region  with different values of exponent (1 $<$ $\beta$ $<$ $\sim$ 3) for different other materials:  in amorphous superconductor Mo$_3$Si (α - Mo$_3$Si) \cite{Clinton}, in indium films \cite{Okuma} and in Fe – or Co- based SC \cite{Wang}$^,$\cite{Hechang}$^-$\cite{Sato}: NaFe$_{1 –x}$Co$_x$As \cite{Wang}, Fe$_{1+y}$(Te$_{1+x}$S$_x$) \cite{Hechang}, Ba(Fe$_{1−x}$Co$_x$)$_2$As$_2$  \cite{Wang2}$^,$\cite{Sato}. 

The aim of our study was to establish the correlation between the longitudinal and the Hall resistivity in the mixed state of electron-doped compounds Nd$_{2-x}$Ce$_x$CuO$_{4+\delta}$ in underdoped region ($x$ = 0.14) and with different degree of nonstoichiometric disorder ($\delta$) near the antiferromagnetic - superconductor boundary. 

\section{Samples and equipment}

We have investigated epitaxial single-crystal films Nd$_{2-x}$Ce$_x$CuO$_{4+\delta}$/SrTiO$_3$ synthesized by pulsed laser deposition \cite{Ivanov} with $x$ = 0.14 (underdoped region) with (001) - orientation (c-axis is perpendicular to the substrate SrTiO$_3$). The films were subjected to a heat treatment (annealing) under different conditions to obtain samples with different contents of oxygen and, consequently, with different degrees of nonstoichiometric disorder. For films with $x$ = 0.14 were obtained two types of samples: optimally annealed in vacuum ($t$ = 60 min, $T$ = 600$^0$C, $p$ =  10$^{-5}$ mmHg) and annealed in vacuum ($t$ = 20, 30, 64 minutes, $T$ = 600$^0$C, $p$ = 10$^{-5}$ mmHg). The thickness of the films was $d$ = 1600-3800 {\AA}.

Consistently annealing Nd$_{2-x}$Ce$_x$CuO$_{4+\delta}$ (NCCO) under different conditions and comparing the results of the  transport properties study \cite{Charikova3,Charikova4} with the results of the structure studies of the oxidized and annealed samples in the neutron diffraction experiments\cite{Schultz}, we have found that the change in the oxygen concentration leads to a change in the degree of nonstoichiometric disorder. Annealing affects the distribution of oxygen in the structure  NCCO, oxygen atoms are redistributed in the positions O (1) and O (2) and removed from the apical position O (3). Annealing in an oxygen-free atmosphere leads to a change in impurity scattering, while having little effect on the concentration of charge carriers \cite{Charikova3}. We will use the concepts of the theory of random two-dimensional systems where the parameter $k_F$$\ell$ serves as the measure of disorder in the system and can be found from the experimental value of $\rho_{xx}$ \cite{Lee}: 

\begin{equation}
k_F\ell = \frac{hc_0}{e^2 \rho_{xx}},\label{eq:2}
\end{equation}

here $c_0$ is the interlayer distance ($c_0$ = 6\AA\ for Nd$_{2-x}$Ce$_x$CuO$_{4+\delta}$) and e is the electron charge.

So the single crystal film optimally annealed in vacuum has $k_F$$\ell$ = 10.8 and the films annealed in vacuum at $t$ = 20; 30, 64 min have disorder parameters  $k_F$$\ell$ = 6.0; 8.8; 2.3 respectively. 

The systematic measurement of the resistivity tensor as a function of the external magnetic field (I$\Vert$ab, B$\Vert$c) in the temperature range $T$ = (0.4 - 40)K were made in a mixed and normal states of electron - doped high-temperature superconductor Nd$_{2-x}$Ce$_x$CuO$_{4+\delta}$ with  $x$ = 0.14 and with varying degrees of nonstoichiometric disorder $\delta$ (different conditions annealing).The resistivity measurements were performed the 4-contact dc-method at the facility PPMS 9 (external magnetic field up to $B$ = 9T, temperature range is $T$ = (1.8 - 40)K in IMP UB RAS. Magnetic field dependencies of the resistivity measurements in the temperature $T$ = (0.4 - 4.2) K were performed in the solenoid "Oxford Instruments" in magnetic fields up to $B$ = 12 T (IMP UB RAS).

\section{Experimental results}
\begin{figure}
\includegraphics{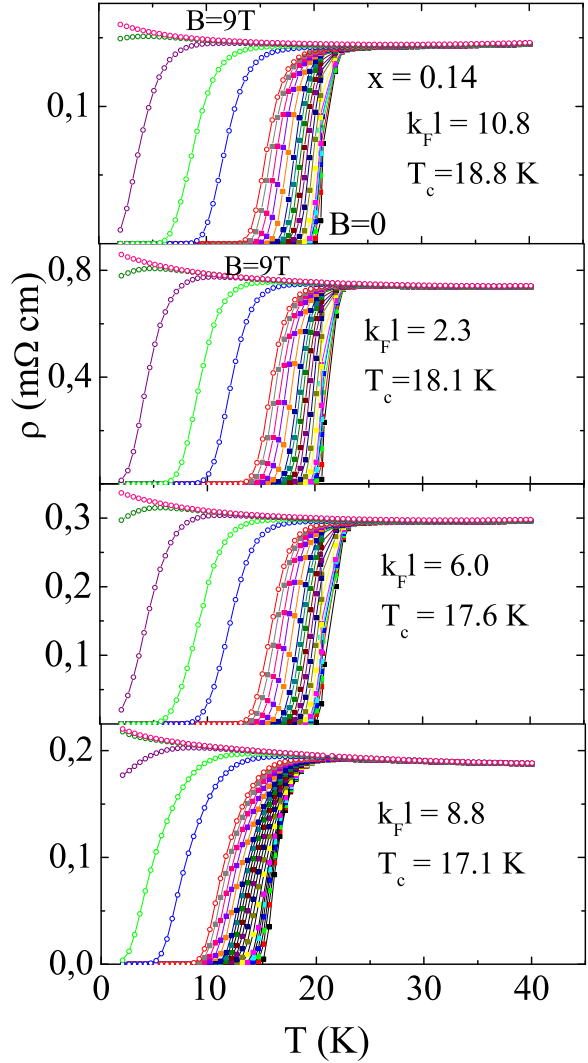}
\caption{\label{fig:Fig1}Temperature dependencies of the longitudinal resistivity $\rho_{xx}$$(T)$ in magnetic field $B$ for single crystal films of underdoped ($x$ = 0.14) Nd$_{2-x}$Ce$_x$CuO$_{4+\delta}$ with different disorder parameters.}
\end{figure}

The in-plane longitudinal $\rho_{xx}$ resistivity were measured as the functions of temperature at $T$=(1.8 - 40)K in perpendicular to the ab-plane magnetic field $B$ up to 9T in electron-doped Nd$_{2-x}$Ce$_x$CuO$_{4+\delta}$/SrTiO$_3$ single crystal films. Figure~\ref{fig:Fig1} shows the temperature dependencies of the longitudinal (in-plane) resistivity $\rho_{xx}$$(T)$ of Nd$_{2-x}$Ce$_x$CuO$_{4+\delta}$/SrTiO$_3$  films with $x$ = 0.14 (underdoped region) and different degree of nonstoichiometric disorder at the increasing of  magnetic fields.  

\begin{figure}
\includegraphics{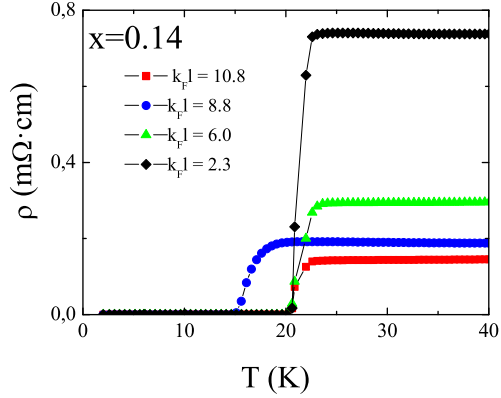}
\caption{\label{fig:Fig2}Temperature dependencies of the longitudinal resistivity $\rho_{xx}$$(T)$ for single crystal films of underdoped ($x$ = 0.14) Nd$_{2-x}$Ce$_x$CuO$_{4+\delta}$ with different disorder parameters at $B$ = 0.}
\end{figure}

Figure~\ref{fig:Fig2} shows $\rho_{xx}(T)$  dependencies for the same films at $B$ = 0 in the temperature range $T$ = (1.8 - 40) K. The deviation from the optimally treatment conditions leads to the increase of nonstoichiometric degree of disorder  from $k_F$$\ell$ = 10.8 for the films optimally annealed in vacuum to $k_F$$\ell$ = 2.3 for the non optimally annealed films  at $T$ = 600$^0$C, $p$ = 10$^{-5}$ mmHg during $t$ =  64 minutes. Temperature dependencies of the resistivity in external magnetic field with  the step $\Delta$B =  0.1 T in the range $B$ = (0 - 3) T and $\Delta$B =  1 T in the range $B$ = (3 - 9) T indicate the parallel shift of the superconducting (SC) transition temperature to the lower temperature region. Any hints on the SC transition disappear above $B$ $\cong$ 7 T (Fig.~\ref{fig:Fig1}).

\begin{figure}
\includegraphics{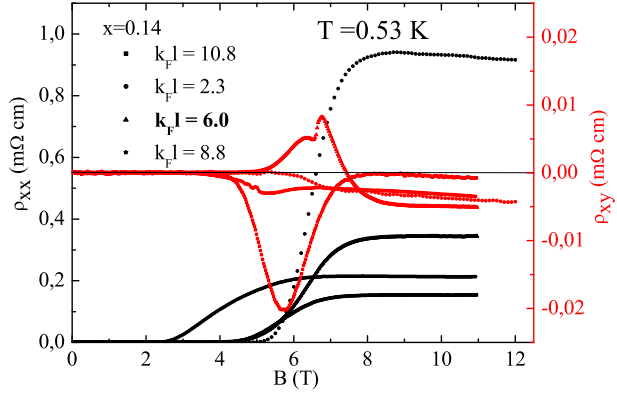}
\caption{\label{fig:Fig3}Magnetic field dependencies of the longitudinal resistivity $\rho_{xx}$$(B)$  and Hall resistivity $\rho_{xy}$$(B)$ for single crystal films of underdoped ($x$ = 0.14) Nd$_{2-x}$Ce$_x$CuO$_{4+\delta}$ with different disorder parameters at $T$ = 0.53 K.}
\end{figure}

\begin{figure}
\includegraphics{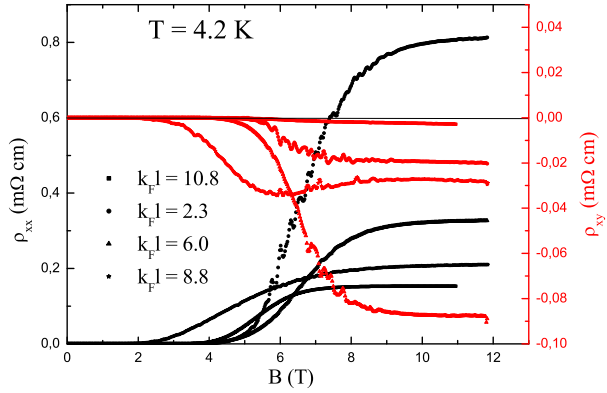}
\caption{\label{fig:Fig4}Magnetic field dependencies of the longitudinal resistivity $\rho_{xx}$$(B)$  and Hall resistivity $\rho_{xy}$$(B)$ for single crystal films of underdoped ($x$ = 0.14) Nd$_{2-x}$Ce$_x$CuO$_{4+\delta}$ with different disorder parameters at $T$ = 4.2 K.}
\end{figure}

The magnetic field dependencies of the longitudinal and Hall resistivities of underdoped ($x$ = 0.14) superconductor Nd$_{2-x}$Ce$_x$CuO$_{4+\delta}$ with varying degrees of nonstoichiometric disorder and hence with different disorder parameters $k_F$$\ell$ at $T$ = 0.53 K and $T$ = 4.2 K are shown in Fig.~\ref{fig:Fig3} and in Fig.~\ref{fig:Fig4} respectively. As one can see the onset of $\rho_{xx}(B)$ and $\rho_{xy}(B)$ occurs at the same magnetic field both for $T$ = 0.53 K and $T$ = 4.2 K. Feature of behavior of $\rho_{xy}(B)$ at $T$ = 0.53 K is the peak observed in the mixed state. Moreover the amplitude of the  peak decreases with the increase of the degree of disorder (disorder parameter decreases)  and changes the sign of the $\rho_{xy}(B)$ peak in the mixed state changes at the minimal annealing time. 

\begin{figure}
\includegraphics{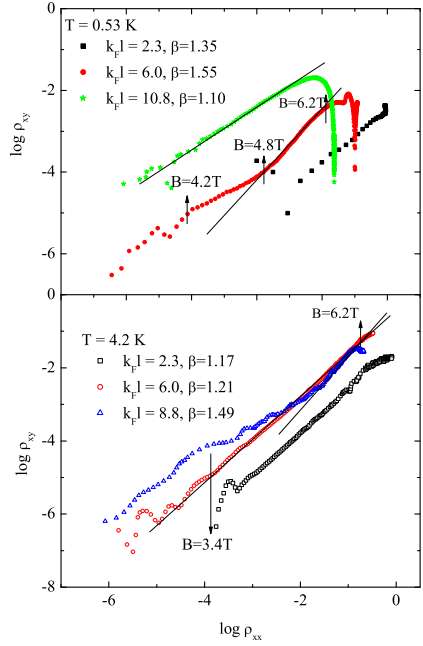}
\caption{\label{fig:Fig5}Log-log plot of the Hall resistivity $\rho_{xy}$ vs longitudinal resistivity $\rho_{xx}$  for underdoped ($x$ = 0.14) Nd$_{2-x}$Ce$_x$CuO$_{4+\delta}$/SrTiO$_3$ films with different disorder parameters at $T$ = 0.53 K (a) and $T$ = 4.2 K (b).}
\end{figure}

In the region of magnetic fields, where the longitudinal and Hall resistivity is vanishingly small it can be detected the appearance of a certain relation between them (Fig.~\ref{fig:Fig5}). The log-log plot reflecting a power-law relationship in the low-resistivity region is presented for Nd$_{2-x}$Ce$_x$CuO$_{4+\delta}$ films with different disorder parameters $k_F$$\ell$ at $T$ = 0.53 K (Fig.~\ref{fig:Fig5}a) and $T$ = 4.2K (Fig.~\ref{fig:Fig5}b). The region of the magnetic fields where we can determine the correlation between the longitudinal and Hall resistivity is located on the first  half of SC transition to the normal state with the increase of the magnetic field $B$ $\approx$ (3 - 6) T regardless of the Hall resistivity sign  in the mixed state.  We can written the ratio as $\rho_{xy}(T,B)$ = $A(T,B)$$\rho_{xx}(T,B)^{\beta}$. The decrease of the disorder degree ($k_F$$\ell$ increases) accompanies the increase of the index $\beta$ both at the $T$ = 0.53 K and $T$ = 4.2 K. 

\begin{figure}
\includegraphics{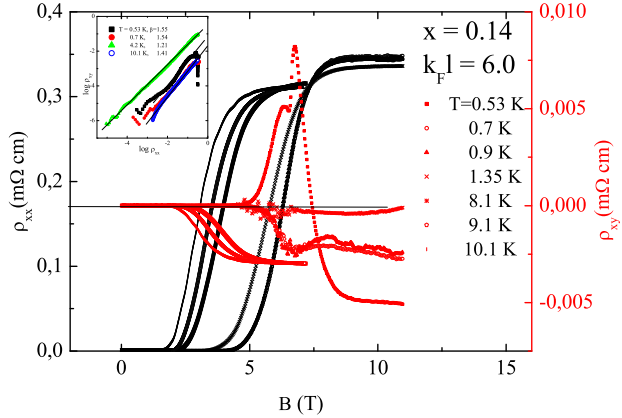}
\caption{\label{fig:Fig6}Magnetic field dependencies of the longitudinal resistivity $\rho_{xx}$$(B)$  and Hall resistivity $\rho_{xy}$$(B)$ for  underdoped Nd$_{2-x}$Ce$_x$CuO$_{4+\delta}$/SrTiO$_3$ films ($k_F$$\ell$ = 6.0) at different temperatures. The insert shows the log-log plot of the Hall resistivity $\rho_{xy}$ vs longitudinal resistivity $\rho_{xx}$  at the sertain temperatures.}
\end{figure}

To find a possible relation of the index $\beta$ with  the temperature, we have made a series of measurements of the longitudinal and Hall resistivity versus magnetic field at a certain temperature at the same time. The results of these measurements for one of the Nd$_{2-x}$Ce$_x$CuO$_{4+\delta}$/SrTiO$_3$ films are shown in Fig.~\ref{fig:Fig6}.  As you can see the index $\beta$ is practically temperature independent from the low temperature ($\beta$ = 1.55) up to the temperature near the SC transition ($\beta$ = 1.41) (see Fig.~\ref{fig:Fig6}, insert).

\section{Discussion}

In a pioneer work on a scaling behavior of the Hall versus longitudinal resistivities \cite{Luo} such a behavior of the Hall resistivity has been attributed to the collective  vortex motion in the presence of pinning.
Very interesting ideas to explain the scaling  of the Hall resistivity have been put forward by Dorsey and Fisher \cite{Dorsey1}. The interrelation  of the Hall and the longitudinal resistivity has been attributed by them to the real scaling regime near the vortex-glass phase transition. 

In a sought-after paper \cite{Vinokur} Vinokur et al.  have calculated the effect of flux pinning on the Hall and longitudinal resistivity and have shown that the pinning straightforwardly gives rise to the scaling relation $\rho_{xy}$ $\propto$ $\rho_{xx}$$^\beta$ with $\beta$ = 2. Such a scaling law arises in a natural way without the hypothesis of the vortex-glass phase transition. The authors argued that a scaling behavior of the Hall resistivity in a mixed state of high-TC superconductors  should be a general feature of any vortex state with disorder-dominated dynamics (vortex liquid, thermally assisted flux flow (TAFF), vortex glass, etc.). 

The theory of Vinokur et al. is based on a force balance equation where the Lorentz force $\bf f$$_L$
acting on a vortex is balanced by the usual frictional force  -$\eta$$\bf v$  and the Hall force $\alpha$$\bf v$$\times$$\bf n$ with $\bf v$ being the average velocity of vortices and $\bf n$ being the unit vector along the
vortex line. The coefficient $\alpha$ is related to the Hall angle by means of tan$\Theta_H$ = $\alpha$/$\theta$.   The explicit expression for $\alpha$  has to be found from microscopic considerations.
On averaging over disorder, thermal fluctuations, and vortex positions one arrives at equation:

\begin{equation}
\eta \bf v + \alpha \bf v \times \bf n =\bf f_{\it L} +\langle F_{pin,\it i} \rangle,
\end{equation}  

where $\bf f_{\it L}$ = ($\Phi_0/c$)$\bf j$ $\times$ $\bf n$, $\bf j$ is the transport current density and  $\Phi_0$ = $\pi$$\hbar$$\it c$/e is the flux quantum, $\langle F_{pin,\it i} \rangle$ is the average pinning force. 

In Ref. \cite{Vinokur} a pinning force have been written in the form:

\begin{equation}
\langle F_{pin,\it i} \rangle = - \gamma (v) \bf v ,   
\end{equation}  

which renormalizes only the frictional term but does not affect the Hall force term. 
The electric field induced by the vortex motion is $\bf E$ = (1/$\it c$)$\bf B$$\times$ $\bf v$ and after solution of Eq. (2) it was obtained:

\begin{equation}
E_x = [B \Phi_0 /c^2 \tilde \gamma (v)]j, E_y = [\alpha B \Phi_0 / c^2 \tilde \gamma^2(v)]j, 
\end{equation}  

where $\tilde \gamma (v)$ = $\gamma (v)$ + $\eta$. It is seen that really pinning renormalizes the friction coefficient $\eta$ only, whereas the Hall "conductivity" $\alpha$ remains unchanged.
From (4) Vinokur et al. received  for resistivity  $\rho$ = $E/j$:

\begin{equation}
\rho_{xy} = A \rho_{xx}^2
\end{equation} 

with
\begin{equation} 
A = \frac{c^2 \alpha}{B \Phi_0}. 
\end{equation} 

Thus, Ref. \cite{Vinokur} demonstrates that pinning does not affect the Hall conductivity and the Hall
resistivity should scale as  $\rho_{xy}$ $\sim$ $\rho^{\beta}_{xx}$ with $\beta$ = 2 for small Hall angles. A scaling with $\beta$ $\approx$ 2 has been really observed in many experiments for p-type HTSes  \cite{Luo}$^-$\cite{Kang5}, for MgB$_2$ \cite{Kang6}$^,$\cite{Soon} and for Fe- and Co - based SC  \cite{Wang}$^,$\cite{Hechang}$^-$\cite{Sato}. The strongest experimental support for the above ideas comes from the finding of Samoilov in Bi-based HTS \cite{Samilov1}  and of Kang et al. \cite{Kang6} and Jung et al. 
\cite{Soon}  in MgB$_2$  which observed a scaling $\rho_{xy}$ $\sim$ $\rho^{\beta}_{xx}$  with $\beta$ = 2.0 $\pm$ 0.1 and a \textit{field independent factor} A.

Comparing (6) to the result of Ref's  \cite{Samilov1}$^,$\cite{Kang6}$^,$\cite{Soon}  one then finds  $\alpha$ $\sim$ $B$ in agreement with the conventional flux motion theory of Bardeen and Stephen (BS) \cite{Bardeen} which predicts that the Hall angle in the mixed state is given by tan$\Theta_H$ = $\omega_c$ $\tau$, where, $\omega_c$ = $eB/mc$  ($e$ and $m$ are the electronic charge and the effective mass, and $\tau$ is the relaxation time in the normal core). In the BS approach the Hall effect is due to the normal carrier Hall effect in the vortex cores, which in turn is proportional to the magnetic field within the core. 

However, the scaling exponent was not always universal and varied from $\beta$ $\approx$ 2 to $\beta$ $\approx$ 1 depending on the applied magnetic field and concentration of defects.
Note that  the other flux flow model of  Nozieres and Vinen (NV) \cite{Nozieres} predicts a constant Hall angle below the upper critical field $H_{c2}$, with the equation tan$\Theta_H$ = $e H_{c2}$ $\tau$/$mc$. As $\rho_{xy}$ = $\rho_{xx}$ $tan\Theta_H$ NV predicts $\beta$ = 1 in a magnetic field-sweep scaling relation.

In their  work \cite{Nozieres} Nozières and  Vinen examined in detail the motion of a flux line in a clean superconductor of extreme type II for very low temperatures. NV suggested that the vortex core is subjected  to a Magnus force identical to that found in liquid helium and that just the Magnus force is an origin of the longitudinal component of the vortex velocity and then of the  Hall voltage
in a mixed state of superconductor. NV have pointed out that at temperatures near $T_c$ and for dirty materials, on which most of the experiments are conducted, some modification of the original model may be required.

The effect of a strong pinning on the Hall effect is not well understood. Wang et al. \cite{Wang1} developed a theory for the Hall effect that includes both pinning and thermal fluctuations based on the normal core model proposed by Bardeen and Stephen. According to Ref.\cite{Wang1} the scaling behavior is explained by taking into account the counter flow in the vortex core due to pinning. As a result, the exponent $\beta$ should change from 2 to 1.5 as the pinning strength increases.
A work of Kopnin and Vinokur \cite{Kopnin1} is a generalization of the work \cite{Vinokur} to the case of a strong pinning. They demonstrate that pinning strongly renormalizes both longitudinal and Hall resistivity in the flux flow regime. Using a simple model for the pinning potential they show that the scaling relation is  violated with an increase in the pinning strength. The Hall resistivity $\rho_{xy}$ scales as $\rho_{xx}^2$ only for a weak pinning. 

In Ref. \cite{Kopnin1} the pinning force after averaging takes the form: 

\begin{equation} 
\langle F_{pin} \rangle = F_1(v) \textbf{v} + F_2(v)[\bf v \times \bf z].
\end{equation} 

where $\bf z$ is the unit vector in the direction of the magnetic field. 
Kopnin and Vinokur \cite{Kopnin1} argue that both coefficients $F_1$ and $F_2$ are generally
nonzero as Eq.(2) does not possess a definite symmetry under the time-reversal transformation $t$ $\to$ -$t$ and $B$ $\to$ -$B$ because of a finite dissipation $\eta$ $\ne$ 0. Thus the pinning force will contain  the dependence both on the direction of the magnetic field and on the direction of the average vortex velocity $\bf v$. 

As a consequence of Eq.(7),  pinning may affect both longitudinal and transverse flux flow conductivity
through the modified force parameters:

\begin{equation} 
\eta \to \eta - F_1 ; \alpha \to \alpha - F_2.
\end{equation} 

It was shown in Ref. \cite{Vinokur}  that for a weak pinning the factor $F_2$ is small: it vanishes faster than $F_1$ as the pinning strength decreases. Thus, for a weak pinning  the Hall conductivity remains almost intact and the scaling $\rho_{xy}$ $\sim$ $\rho_{xx}^2$ is restored. But  in general for a strong enough pinning  should  be no  universal scaling of the Hall and longitudinal resistivities.

The $d$-wave superconductivity adds new aspects to the problem under discussion \cite{Kopnin2}: an existence of the gap nodes  lead to a peculiar scaling law for thermodynamic and kinetic properties of $d$-wave superconductors at low $T$ $\ll$ $T_c$ in small magnetic fields $B$ $\ll$ $B_{c2}$. Thus, it results in a new scaling law for the Ohmic and Hall conductivities in a mixed state of  $d$-wave superconductor.
Accoding to Ref.\cite{Kopnin2},  the Hall angles for clean $d$-wave superconductors  in the low $T$ and $B$ regimes should have a universal value independent of the mean free paths ($\ell$) of electrons and holes. Their main conclusion for the relation between $\rho_{xy}$ and $\rho_{xx}$ is given by

\begin{equation} 
\rho_{xy} = (\pi/2) \frac{(n_e - n_h)}{(n_e + n_h)} \rho_{xx} ,
\end{equation} 

where $n_e$ and $n_h$ are the electron-like and hole-like charge carrier density, respectively. 

In TAFF region, since $n_e$ and $n_h$ are slowly varying functions while $\rho_{xx}$ is an exponential function of $T$, Eq.(9) leads to a universal scaling with $\beta$ = 1 independent of the mean free path $\ell$ and $B$.

Kang et al. have studied the Hall scaling relation $\rho_{xy}(T)$ = A $\rho^{\beta}_{xx}(T)$ for the mixed-state Hall effects of HgBa$_2$CaCu$_2$O$_{6+\delta}$ thin films before and after irradiation by high-energy Xe ions and observed that $\beta$ increases from 1.0 $\pm$ 0.1 to 2.0 $\pm$ 0.1 as a perpendicular magnetic field increases from 0 to 8 T.  They believe that the observation of $\beta$ = 1 at relatively low $H$  in Hg-1212 film with columnar defects is consistent with theory of Ref.\cite{Kopnin2} based on $d$-wave superconductivity.

According to Ref.\cite{Kopnin2}, however, the sample purity needed for observation of the universal behavior of Eq. (9) is rather high; it requires $\ell$/$\xi_0$ $\gg$ 1 ($\xi_0$ being the SC coherence length) depending on the
particular high-$T_c$ superconductor.

While  it is generally accepted that p- type cuprate superconductors are d-wave superconductors \cite{Izyumov}$^,$\cite{Moriya}, for n - type HTS systems the situation is not so definite, and different experimental data indicate the benefit of  $d$-type symmetry or of $s$-type anisotropic symmetry \cite{Armitage}$^,$\cite{Charikova5}. It should not be excluded that the observation of $\beta$ = 1.10 $\pm$ 0.05 for the optimally annealed samples ($k_F$$\ell$ = 10.8) at the lowest temperature ($T$ = 0.53 K) in all scaling field interval  at  3.2 T $<$ $B$ $<$ 5.5 T  (see Fig.~\ref{fig:Fig5}a) is a manifestation of anisotropic $s$ - wave or $d$-wave pairing symmetry.

Generally, the values of $\beta$ = 1.5 $\pm$ 0.4 observed in the investigated samples substantially less than the universal value $\beta$ = 2 predicted by a widely used phenomenological flux-flow model of Vinokur et al. \cite{Vinokur}.
It is well known (see \cite{Charikova4} and references therein) that the compound Nd$_{2-x}$Ce$_x$CuO$_{4+\delta}$ is characterized by the ability to reversibly absorb and release oxygen and in the samples subjected to different heat treatment to obtain a different content of oxygen ($\delta$), there are varying degrees of nonstoichiometric disorder.
 Furthermore, an underdoped compound Nd$_{2-x}$Ce$_x$CuO$_{4+\delta}$ with x = 0.14 is on a border of the antiferromagnetic and superconducting phases in electron-doped cuprate system \cite{Armitage} that is near the quantum phase transition antiferromagnetic - superconductor. Strong fluctuations developing in the vicinity of the continuous quantum phase transition can result in substantial macroscopic inhomogeneity of the system.

Thus, due to the considerably large concentration of defects
in the investigated system we most likely are in conditions of applicability of the flux-flow models with the strong pinning strength \cite{Wang1}$^,$\cite{Kopnin1}. Wang et al.\cite{Wang1} claimed that $\beta$ could change from 2 to 1.5 as the pinning strength increased, which agreed with the results reported for YBCO crystals \cite{Kang3} and HgBa$_2$CaCu$_2$O$_6$ films \cite{Kang5}. On the other hand, in the model of  Kopnin and Vinokur \cite{Kopnin1}, in which  pinning may affect both longitudinal and transverse flux-flow conductivity, it should in general be no universal $\beta$ = 2  scaling of the Hall and longitudinal resistivities  for a strong enough pinning.

\section{Conclusions}
  
The systematic magnetic-field dependence investigations of the longitudinal and Hall resistance of the electron-doped compounds Nd$_{2-x}$Ce$_x$CuO$_{4+\delta}$ in underdoped region with $x$ = 0.14 and with varying degrees of disorder ($\delta$) were held. It was established experimentally that the correlation between the longitudinal electrical resistivity and the Hall resistivity can be analyzed on the basis of scaling relationships: $\rho_{xy}$(B)$\sim$ $ \rho_{xx}$$^\beta$(B). For a set of single-crystal films of Nd$_{2-x}$Ce$_x$CuO$_{4+\delta}$/SrTiO$_3$ investigated by us the value $\beta$ = 1.5 $\pm$ 0.4 is found which is substantially less than the theoretical value $\beta$ = 2 predicted by a widely used phenomenological flux-flow model of Vinokur et al.\cite{Vinokur}. The observed feature in the electron-doped HTS system as well as in a number of hole-doped HTS can be explained by a rather strong pinning due to an essential degree of the sample inhomogeneity. 
Such an inhomogeneity may be associated both with the nonstoichiometric disorder inherent to Nd$_{2-x}$Ce$_x$CuO$_{4+\delta}$  system and with the particularity of $x$ = 0.14 underdoped compound disposed in a vicinity of  the antiferromagnetic-superconductor transition.

\begin{acknowledgments}
The research was carried within the state assignment of theme “Electron” (No 01201463326), supported in part by the Program of fundamental research of the Ural Division of Russian Academy of Sciences (RAS) "Quantum macrophysics and nonlinear dynamics" (Рroject № 15-8-2-6) with partial support of RFBR (grant N 15-02-02270).
\end{acknowledgments}

\end{document}